\documentstyle[prl,aps,multicol,psfig]{revtex}
\begin{document}
\draft

\title{From the Chern-Simons theory for the fractional
quantum Hall effect to the Luttinger model of its edges}

\author{Dror Orgad}

\address{Department of Condensed Matter Physics, The Weizmann Institute of Science,
Rehovot 76100, Israel}

\date{\today}

\maketitle

\begin{abstract}
The chiral Luttinger model for the edges of the fractional quantum Hall effect is obtained
as the low energy limit of the Chern-Simons theory for the two dimensional system. In 
particular we recover the Kac-Moody algebra for the creation and annihilation operators of 
the edge density waves and the bosonization formula for the electronic operator at the edge.
\end{abstract}


\begin{multicols}{2}
The fractional quantum Hall effect (FQHE) is a manifestation of the formation of an 
incompressible state in a two dimensional electron system at discrete, density dependent
values of the magnetic field. The major step towards a theoretical understanding of
this phenomenon was taken by Laughlin \cite{Laughlin83} who proposed a variational many 
body wavefunction to describe a correlated, incompressible electron liquid at filling 
factors $\nu = 1/\tilde\phi$ for odd $\tilde\phi$. The microscopic wavefunction approach
\cite{Laughlin83,Jain89} was subsequently augmented by an effective field theory of the 
Chern-Simons type. As it turned out both the bosonic \cite{Zhang89} and fermionic 
\cite{Lopez91} versions of the theory proved to be fruitful in explaining different aspects 
of the effect. While the incompressibility of the FQHE implies a gap for the excitations in 
the bulk of the system the edges support gapless modes that control the low energy physics
of the problem \cite{Halperin82}. Wen pioneered the use of the chiral Luttinger 
model to describe these edge excitations and indicated, among others 
\cite{Wen90,Wen91,Wen92,Kane92,Oreg96}, various non Fermi liquid effects which arise from this 
description. While plausible arguments, supported by recent numerical calculations 
\cite{Palacios96}, have been put forward in favor of the applicability of the Luttinger
model to the edges of the FQHE \cite{Wen90,Wen91,Wen92,Stone94}, a microscopic justification is 
still lacking. It is the purpose of the following Letter to supply such a justification 
for the fractions of the form $\nu = 1/\tilde\phi$. In doing so we reveal an intimate 
relation between the bosonic Chern-Simons theory for the two dimensional system and the one 
dimensional Luttinger model of its edges. The latter is obtained as the low energy limit 
of the former. In particular we recover the Kac-Moody algebra for the creation and 
annihilation operators of the edge density waves and the bosonization formula that enables 
one to express the electronic operator at the edge in terms of them.

The central idea of the Chern-Simons theory is the statistical transmutation of the 
electronic field operators $\psi({\bf r})$ via an operator valued transformation
\begin{equation}
\label{cstransform}
\phi({\bf r})=e^{i\Lambda({\bf r})}\psi({\bf r}) \;,\;\;\;\;\;
\Lambda({\bf r})=\int d^2r' f({\bf r-r'})\rho({\bf r'}) \; ,
\end{equation}
where $\rho({\bf r})=\psi^{\dagger}({\bf r})\psi({\bf r})=\phi^{\dagger}({\bf r})
\phi({\bf r})$. We will be concerned with transformations that lead to bosonic $\phi({\bf r})$ 
and $\phi^{\dagger}({\bf r})$. It can be easily checked using the Baker-Hausdorff formula that 
this is the case when $f({\bf r-r'})=f({\bf r'-r})+\tilde\phi\pi$ for odd $\tilde\phi$.
The Chern-Simons transformation also introduces additional terms in the equation of motion 
of $\phi({\bf r})$ which couple to it as vector and scalar potentials \cite{Lerda92}
\begin{eqnarray}
\label{axy}
&& {\bf a}({\bf r})=-{\hbar c\over{e}}\int d^2r' {\bf F}({\bf r-r'})\rho({\bf r'}) \\
\label{a0}
&& a_{0}({\bf r})=-{\hbar\over{e}}\int d^2r' {\bf F}({\bf r-r'})\rho({\bf r'})\mbox
{\boldmath $v$}({\bf r'})
\end{eqnarray}
where ${\bf F}({\bf r-r'})=\nabla_{r} f({\bf r-r'})$ and $\mbox{\boldmath $v$}({\bf r})$ 
is the velocity operator. These "statistical gauge fields" constitute the main motivation 
for taking step (\ref{cstransform}). For appropriate electronic densities they cancel on
the average the external fields thus removing the degeneracy of the ground state and allowing
for perturbative treatment of the interactions. In the Lagrangian formalism the equivalent 
bosonic system is described by the Chern-Simons action \cite{Zhang89}. Using the polar 
decomposition of the bosonic field $\phi=\sqrt{\rho}e^{i \theta}$ and the velocity fields 
\begin{equation}
\label{vdef}
v_i={\hbar\over{m}}\partial_i\theta+{e\over {mc}}(a_i+A_i) \; ,
\end{equation}  
with $\bf{A}$  taken such as to produce a constant magnetic field along $z$, one obtains 
the hydrodynamic form of the Lagrangian density which is more convenient for our purpose, 
\begin{eqnarray}
\label{lagrangian}
\nonumber \cal L&=& \rho(A_0+a_0-\partial_t\theta)-{1\over 2}\rho v_i^2-{1\over 8}
{{(\partial_i\rho)^2}\over{\rho}}  \\
&-&{1\over{4\pi}}\int d^2r' \rho({\bf r})U({\bf r-r'}) \rho({\bf r'}) \\
\nonumber &+&{\epsilon^{ij}\over{2\tilde\phi}}(v_i\partial_t v_j -2v_i\partial_t\partial_j
\theta) + {a_0\over{\tilde\phi}}(\epsilon^{ij}\partial_i v_j -1) \; .
\end{eqnarray}
Henceforth, unless otherwise specified, length is measured in units of the magnetic 
length $l=\sqrt{\hbar c/eB}$, time in units of inverse cyclotron frequency 
$\omega_c=eB/mc$, energy is normalized by $\hbar\omega_c$ and the density by the Landau level
degeneracy $\rho_0=1/2\pi l^2$.
On deriving (\ref{lagrangian}) several surface terms were neglected as they can be shown 
to vanish for the configurations that will be of interest to us. Minimizing this action results
in the mean-field equations which are the starting point of our analysis
\begin{mathletters}
\begin{eqnarray}
\label{bfield}
&&\epsilon_{ij}\partial_i v_j=1-\tilde\phi\rho\\
\label{efield}
&&\partial_t\partial_i\theta-\partial_t v_i-\partial_i a_0=\tilde\phi\epsilon_{ij}\rho v_j\\
\label{cont}
&&\partial_t\rho=-\partial_i(\rho v_i)\\
\label{sch}
\nonumber
&&\partial_t\theta=-{1\over 2}v_i^2+{1\over 2}{\partial_i^2\sqrt{\rho}\over 
\sqrt{\rho}}+a_0+A_0 \\
&&\hspace{0.96cm} -{1\over{2\pi}}\int  U({\bf r-r'})\rho({\bf r}')d^2{\bf r}'\; . 
\end{eqnarray}
\end{mathletters}

We will consider a semi infinite system defined by an infinitely high wall situated at 
$x\leq 0$ and obeying periodic boundary conditions over length $L$ in the $y$ direction.
Our plan is to obtain the edge excitations as the random phase approximation (RPA) modes
of the theory, i.e., the eigenmodes of the above equations linearized around a static 
solution. This step was carried out in \cite{Orgad96} and its essentials are reproduced below
for the sake of completeness of presentation. Once the gapless modes are at hand we will 
use them to expand the deviations of the various fields from their average values, namely the 
static solution. The expansion coefficients will become, once quantized, the density 
operators of the Luttinger model. Their commutation relations and Hamiltonian will be 
deduced from the quadratic Lagrangian obtained by expanding (\ref{lagrangian}) to second 
order in the deviations. Finally by inverting the transformation (\ref{cstransform}) we will 
recover the bosonization formula. 

\begin{figure}
\narrowtext
      \centerline{\psfig{figure=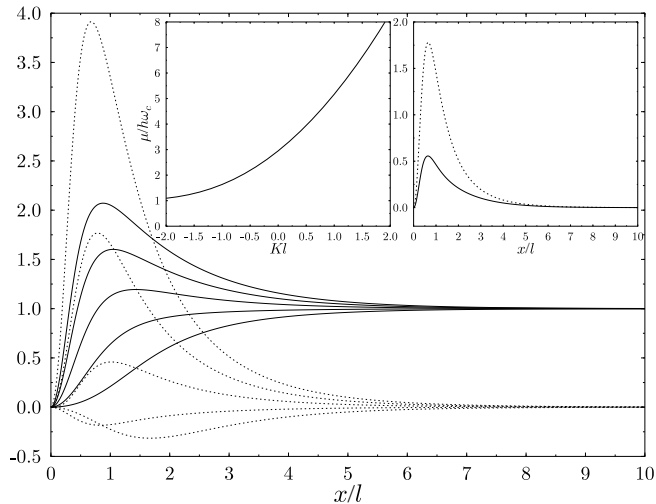,angle=270,width=3.375in}}
\caption{Density in units of $\bar\rho$ (solid lines) and current density in units of 
$\bar\rho\omega_c l$ (dashed lines) for the case of short range interaction $U=(2\pi\hbar^2
/m)\delta({\bf r-r'})$. The highest curve in each set corresponds to $K=2$. Consecutive curves 
differ by $\Delta K=-1$. The right inset depicts the density and current density profiles of the 
translation mode corresponding to the $K=1$ solution while the left one displays $\mu(K)$. 
For more details, especially on the differences between the 
cases of short range and Coulomb interactions, see Ref. \protect\cite{Orgad96}.}
\label{srsol}
\end{figure}

The translational symmetry of the system along the $y$ direction is in conflict with the 
canonical choice 
\begin{equation}
\label{f}
f({\bf r-r'})=\tilde\phi\,{\rm tg^{-1}}\left({{y-y'}\over{x-x'}}\right)
\end{equation}
and the "symmetric gauge" induced by ${\bf F}({\bf r-r'})=\lim_{\varepsilon\rightarrow 0^{+}}\,
\tilde\phi\hat{\bf z}\times ({\bf r-r'})/(|{\bf r-r'}|^2+\varepsilon)$. Note that $f({\bf r-r'})$
is a multi-valued function and we pick its branch cut along the positive $x-x'$ axis. At this 
point we take advantage of the fact that by adding to $f({\bf r-r'})$ a function $g({\bf r-r'})$ 
satisfying $g({\bf r-r'})=g({\bf r'-r})$ we do not affect the statistics of 
$\phi({\bf r})$. Such a change on the operatorial level corresponds to a regular gauge transformation 
in the Lagrangian formalism. We choose 
\begin{equation}
\label{g}
g({\bf r-r'})=\lim_{\varepsilon\rightarrow 0^{+}}\tilde\phi{{y-y'}\over{\sqrt{(y-y')^2+
\varepsilon}}}{\rm tg}^{-1}\!\left[{{x-x'}\over{\sqrt{(y-y')^2+\varepsilon}}}\right]
\end{equation}
which is both single-valued and symmetric under the 
exchange of ${\bf r}$ and ${\bf r'}$. As a result we find that $(f+g)({\bf r-r'})=\pi\tilde
\phi[1-\epsilon(y-y')]$ with a branch cut along the positive $x-x'$ axis and $\nabla_r (f+g)=
[0,2\pi\tilde\phi\epsilon(x-x')\delta(y-y')]$ which in turn implies the "Landau gauge" 
for the statistical potentials  
\begin{eqnarray}
\label{landauaxy}
&& {\bf a}({\bf r})=\left[0\;,\; -\tilde\phi\int d^2r'\epsilon(x-x')\delta(y-y')
\rho({\bf r'})\right] \\
\label{landaua0}
&& a_{0}({\bf r})=-\tilde\phi\int d^2r' \epsilon(x-x')\delta(y-y')\rho({\bf r'})
v_y({\bf r'})
\end{eqnarray}
with $\epsilon(x)=\Theta(x)-1/2$ where $\Theta(x)$ is the step function. We will find it 
more convenient, however, to obtain the static solution and the edge modes in a slightly 
different gauge where we replace $\epsilon(x-x')$ in (\ref{landauaxy},
\ref{landaua0}) by $\Theta(x-x')$. Utilizing this gauge we look for a static solution in 
the form 
\begin{equation}
\label{sol}
\theta=Ky-\mu(K)t \;,\;\;\;\;\; \rho = \rho(x) \; .
\end{equation}
Due to the infinite confining potential we set the density to zero at the wall. For the external
vector potential we use the gauge ${\bf A}=[0,Bx]$ which gives a velocity field 
of the form $v_x=0$ and $v_y=v_y(x)$. Inserting the above ansatz into 
(\ref{bfield})-(\ref{sch}) one finds a set of coupled equations for $\rho(x)$, $v_y(x)$ and 
$a_0(x)$ that was solved numerically. For a fixed value of $K$ the value of $\mu$ was determined
by requiring that $\rho$ approaches its bulk density $\bar\rho=\rho_0/\tilde\phi$ far from the
wall. Under this condition $\mu$ is the energy which is needed in order to add a particle to the
edge. Representative examples of the density and current density profiles for solutions
(\ref{sol}) in the case of short range interactions, along with $\mu(K)$, are shown in Fig.\ 
\ref{srsol}. One finds a one parameter family of static solutions depending on $K$ and differing 
by the density of particles at the edge. Changing the value of $K$ translates the condensate 
$\phi$ in the $x$ direction. Using $v_y(0)=K$ and the fact that the velocity falls to zero in 
the bulk we can integrate (\ref{bfield}) to find that $K/\tilde\phi$ is the excess charge per 
unit length along the edge relative to a step-like constant density profile.

Next we proceed to obtain the edge modes as solutions to the RPA equations, namely the 
linearized version of (\ref{bfield})-(\ref{sch}) around one of the solutions (\ref{sol})
\begin{mathletters}
\begin{eqnarray}
\label{lbfield}
&&\epsilon_{ij}\partial_i \delta v_j=-\tilde\phi\delta\rho\\
\label{lefield}
&&\partial_t\partial_i\delta\theta-\partial_t\delta v_i-\partial_i\delta a_0=
\tilde\phi\epsilon_{iy}v_y\delta\rho+\tilde\phi\epsilon_{ij}\rho\delta v_j\\
\label{lcont}
&&\partial_t\delta\rho=-v_y\partial_y\delta\rho-\partial_i(\rho\delta v_i)\\
\label{lsch}
\nonumber
&&\partial_t\delta\theta=-v_y\delta v_y+p(\rho,\delta\rho)+{\partial_y^2\delta
\rho\over{4\rho}}+\delta a_0 \\
&&\hspace{1.12cm}-{1\over{2\pi}}\int  U({\bf r-r'})\delta\rho({\bf r}')d^2{\bf r}'\; ,
\end{eqnarray}
\end{mathletters}
where $p$ denotes  the part of the linearized  "quantum pressure" term, i.e. the
second term on the r.h.s. of  (\ref{sch}), containing $x$ derivatives. The derivation of the
edge excitation is facilitated by the observation that the derivative of the static solution
with respect to $K$ constitute a solution to (\ref{lbfield}-\ref{lsch}). This is the translation 
mode (along the $x$ direction) of the system and it describes the addition or removal of charge 
from the vicinity of the wall. {\it Apriory} the amount of charge carried by this solution is 
arbitrary since, due to the linearity of (\ref{lbfield}-\ref{lsch}), it is determined only up to 
a multiplicative constant. We will find, however, at the end of the discussion, that the periodic 
boundary conditions impose integer values for this charge in units of $1/\tilde\phi$. We thus 
interpret the translation mode as the state of a smeared quasi-particle along the edge. For a 
representative example of these solutions see Fig.\ \ref{srsol}.

The gapless branch is obtained by modulating the translation mode in the $y$ 
direction. Concentrating on the long-wavelength limit we find to lowest order in $k$
\begin{eqnarray}
\label{lsol}
\nonumber \delta\rho&=&\partial_{\scriptscriptstyle K}\rho(x)e^{-i(ky-\omega t)}\\
\nonumber \delta v_x&=&{i\over{\tilde\phi\rho(x)}}[\omega\partial_{\scriptscriptstyle K} v_y(x)
                       -k\partial_{\scriptscriptstyle K} a_0(x)-\omega]e^{-i(ky-\omega t)}\\
          \delta v_y&=&\partial_{\scriptscriptstyle K}v_y(x)e^{-i(ky-\omega t)}\\
\nonumber \delta a_0&=&\partial_{\scriptscriptstyle K}a_0(x)e^{-i(ky-\omega t)}\\
\nonumber \delta\theta&=&{i\over k}e^{-i(ky-\omega t)} \; .
\end{eqnarray}
The real and imaginary part of (\ref{lsol}) constitute , to first order in $k$, independent 
solutions of (\ref{lbfield}-\ref{lsch}), provided the dispersion relation 
$\omega=\omega(k)$ is taken to be 
\end{multicols}
\widetext

\noindent
\setlength{\unitlength}{1in}
\begin{picture}(3.375,0)
  \put(0,0){\line(1,0){3.375}}
  \put(3.375,0){\line(0,1){0.08}}
\end{picture}
\begin{equation}
\label{ldisp}
{\omega(k)\over k}={\partial\mu\over{\partial K}}+\lim_{k\rightarrow 0} {1\over{2\pi}}\int dx' 
\partial_{\scriptscriptstyle K}\rho(x')\left[\tilde U(x-x',k)-\tilde U(x-x',0)\right] \; ,
\end{equation}
\hfill
\begin{picture}(3.375,0)
  \put(0,0){\line(1,0){3.375}}
  \put(0,0){\line(0,-1){0.08}}
\end{picture}

\begin{multicols}{2}
\noindent
where $\tilde U(x,k)=\int dy U(x,y)e^{iky}$. For short range interaction the last 
term in (\ref{ldisp}) vanishes while for Coulomb interaction it is proportional to $-\ln (k)$ 
minus the electrostatic charging energy of the edge \cite{Orgad96}.

The solutions of the time independent linear eigenvalue problem defined by (\ref{lbfield}-
\ref{lsch}), i.e. the same equations but with time derivatives replaced by $i\omega$, constitute 
a complete set which one can use to expand the deviations of the various fields from their 
static average configurations. We will, however, restrict the sum over eigenmodes to the 
gapless edge excitations found above. In making this approximation we neglect other modes such 
as bulk magneto-plasmons and magneto-rotons \cite{Girvin86,Lee91} and their scattering states 
off the wall. We also do not consider the effects of quasi-particles and quasi-holes in the bulk. 
All of these excitations involve an energy gap and are thus expected to be irrelevant in the low 
energy limit of the theory. At this stage we transform back to the original gauge 
(\ref{landauaxy},\ref{landaua0}). As a result $\theta$ changes to $[K-\tilde\phi/2\int 
dx \rho(x)]y - [\mu-\tilde\phi/2\int dx \rho v_y(x)]t$ and the expansion of the deviation 
fields in the edge modes reads
\begin{eqnarray}
\label{expansion}
\nonumber \delta\rho&=&{2\pi\tilde\phi\over L}\delta\rho(0)\partial_{\scriptscriptstyle K}\rho
+{2\pi\tilde\phi\over L}\sum_{k\neq 0}\delta\rho(k)\partial_{\scriptscriptstyle K}\rho e^{-iky}\\
\nonumber \delta v_x&=&{2\pi i\over{L\rho}}\sum_{k\neq 0}\delta\rho(k)[\omega\partial_
{\scriptscriptstyle K} v_y-k\partial_{\scriptscriptstyle K} a_0-\omega]e^{-iky}\\
          \delta v_y&=&{2\pi\tilde\phi\over L}\delta\rho(0)\partial_{\scriptscriptstyle K}v_y
+{2\pi\tilde\phi\over L}\sum_{k\neq 0}\delta\rho(k)\partial_{\scriptscriptstyle K}v_y e^{-iky}\\
\nonumber \delta a_0&=&{2\pi\tilde\phi\over L}\delta\rho(0)\left[\partial_{\scriptscriptstyle K}
a_0+{1\over 2}{{\partial\mu}\over{\partial K}}\right]\\
\nonumber &+&{2\pi\tilde\phi\over L}\sum_{k\neq 0}\delta\rho(k)\left[\partial_
{\scriptscriptstyle K} a_0 +{1\over 2} {\omega\over k}\right]e^{-iky}\\
\nonumber \delta\theta&=&{\pi\tilde\phi\over L}\delta\rho(0)y+{\pi\tilde\phi\over L}
\sum_{k\neq 0}{i\over k}\delta\rho(k)e^{-iky} \; .
\end{eqnarray}
The factor $2\pi\tilde\phi/L$ in (\ref{expansion}) was introduced, in connection with the fact
that $\int dx \partial_{\scriptscriptstyle K}\rho(x) =1/\tilde\phi$, in order to make the 
configurations carry a unit charge (in dimensionful units) as we are interested to describe 
an {\it electron} at the edge. The expansion coefficient $\delta\rho(0)$ will become, upon 
quantization, the number operator that measures the number of excess electrons on the edge 
relative to the static ground state configuration around which we expand \cite{Haldane81}. 
It is time independent since we are interested in the case where the amount of particles is 
fixed. The rest of the expansion coefficients $\delta\rho(k)$ are the new dynamical degrees of 
freedom and will play the role of the creation and annihilation operators of the edge density 
waves. Since the fields are real they obey $\delta\rho(k)=\delta\rho^{*}(-k)$.
The resulting effective low energy theory is described by the quadratic Lagrangian which 
is obtained by plugging (\ref{expansion}) into (\ref{lagrangian}) and keeping terms to lowest 
order in $k$. Here and in the remaining part of the Letter we restore the units of dimensions.
\begin{eqnarray}
\label{quadlagrangian}
        L^{(2)}&=&-{{2\pi\tilde\phi\hbar}\over L}\sum_{k>0}{i\over k}\partial_t\delta\rho(k)
\delta\rho(-k) \\
\nonumber &-&{{\pi\tilde\phi}\over L}{{\partial\mu}\over{\partial K}}\delta\rho(0)
\delta\rho(0)-{{2\pi\tilde\phi\hbar}\over L}\sum_{k>0}v(k)\delta\rho(k)\delta\rho(-k) \;,
\end{eqnarray}
where $v(k)=\omega(k)/k$. The last two terms of (\ref{quadlagrangian}) are (up to a minus 
sign) the Hamiltonian of the reduced problem. They coincide with the expression for the 
Hamiltonian of the chiral Luttinger model \cite{Wen92,Haldane81}. The first of these terms 
corresponds to the change of energy, relative to the ground state energy, due to excess charge 
on the edge. The second term describes the energy of the gapless excitations. 
The effects of intra-edge interactions are included in this Hamiltonian through the
coefficient $\partial\mu/\partial K$ and the dispersion relation $v(k)$. The commutation 
relations of the density operators are inferred from the symplectic part of $L^{(2)}$, that is 
the first term in (\ref{quadlagrangian}). We find that the conjugate momentum to $\delta\rho(k>0)$
is $-(2\pi\hbar\tilde\phi i/kL)\delta\rho(-k)$ thus giving, as anticipated, the 
algebra of the chiral Luttinger model 
\begin{equation}
\label{commrel}
[\delta\rho(-p),\delta\rho(k)]={{kL}\over{2\pi\tilde\phi}}\delta_{k,p} \;\;\;\;\;\;\; 
{\rm for}\; k>0 \; .
\end{equation} 
Note the factor $1/\tilde\phi$ in (\ref{commrel}) which is responsible for the non Fermi liquid 
behavior of the theory \cite{Wen90,Wen91,Wen92}.
This algebra should be supplemented by a ladder operator $U$ that raises the fermion charge 
at the edge in $1/\tilde\phi$ steps \cite{Haldane81}. Both $U$ and the number operator 
$\delta\rho(0)$ commute with the rest of the operators.

Within the low energy approximation and to lowest order in $k$ we find that the bosonic 
operators, in the gauge defined by (\ref{landauaxy},\ref{landaua0}), are given by $\phi({\bf r})=
\sqrt{\rho(x)}\exp{[i(\theta+\delta\theta)(y)]}$. This expression can be used together with 
Eqs.\ (\ref{f}) and (\ref{g}) in order to invert the Chern-Simons 
transformation (\ref{cstransform}) to obtain an approximate form for the fermionic operators. 
The result consists, presumably, of the low energy edge components of the exact $\psi({\bf r})$. 
Finally we obtain a one dimensional operator by projecting the two dimensional 
$\psi({\bf r})$ on the edge modes through an integration over the $x$ direction with a weight given 
by the profile of their wavefunctions $\partial_{\scriptscriptstyle K}\rho(x)/\sqrt
{\rho(x)}$. The emerging expression is the bosonization formula for the electronic operator of 
the chiral Luttinger model \cite{Wen92,Haldane81} 
\begin{eqnarray}
\label{bosonization}
          \psi(y)&=&L^{-1/2}\exp[i\varphi(y)]\\
\nonumber \varphi(y)&=&Ky+{2\pi\tilde\phi\over L}\delta\rho(0)y+{2\pi\tilde\phi\over L}
\sum_{k\neq 0}{i\over k}\delta\rho(k)e^{-iky} \; .
\end{eqnarray}
In the calculation leading to (\ref{bosonization}) $y$ independent terms in $\Lambda({\bf r})$
were ignored. These terms do not arise in the case of an infinite system  
if one assumes that the density vanishes at infinity. They do occur, however, for the case of a 
periodic system which we consider here, and they have to be omitted  
in order to make $\psi({\bf r})$ obey the correct anti-commutation relations. 
It is interesting that $\varphi(y)$ is composed, in equal parts, of contributions from 
the Chern-Simon phase $\Lambda$ and from the dynamical phase $\delta\theta$ of the bosonic field. 
We conclude by noting that the requirement that (\ref{bosonization}) should obey periodic
boundary conditions imposes a discrete spectrum for the number operator 
$\delta\rho(0)$, in units of $1/\tilde\phi$. This observation is in accord with our earlier 
identification of the translation mode as the state of a quasi-particle at the edge.

It is a pleasure to thank S. Levit, A. M. Finkel'stein and Y. Oreg for many useful discussions.

\end{multicols}
\end{document}